\pdfoutput=1  

\documentclass[aps,prd,superscriptaddress,floatfix,nofootinbib,preprintnumbers,
eqsecnum,twocolumn
]{revtex4-1}

\usepackage{amsmath,float,graphicx,placeins,amssymb,multirow,pgfplots,pgfplotstable}
\usepackage{tikz}
\usepackage{soul}
\setstcolor{red}
\usepackage{tabularray}

\usepackage{lipsum} 
\usepackage{comment}
\usepackage{enumerate,slashed,cancel,comment,color,bbm,graphicx,rotating,placeins,mathtools,multirow,hhline}
\usepackage[normalem]{ulem}
\usepackage{array}
\usepackage{enumitem}
\usepackage{hyperref}
\usepackage{color}
\usepackage{bm}
 \hypersetup
 { 
   colorlinks,
   linkcolor={blue!80!black},
   citecolor={blue!70!black},
   urlcolor={blue!70!black}
 }

\usepackage{diagbox}
\usepackage{tensor}
\usepackage{subfig}
\usepackage{physics}

\usetikzlibrary{shapes.geometric,arrows}
\graphicspath{}

\definecolor{oxfordblue}{rgb}{0.0, 0.13, 0.28}
\definecolor{cambridgeblue}{rgb}{0.64, 0.76, 0.68}

\newcommand{\cG}{\mathcal{G}}

\newcommand{\uv}{\mathbf{e}}

\newcommand{\bq}{\mathbf{q}}

\newcommand{\IP}{\mathbb{P}}

\newcommand{\varstr}[2]{\vrule height #1 depth #2 width0pt}

\newcommand{\cicy}[2]{\begin{matrix} #1\end{matrix}\!\left[\begin{matrix}#2 \end{matrix}\right]}

\begin{document}
\date{Last updated: \today}

\title{Reproducing Standard Model Fermion Masses and Mixing in String Theory:\\ A Heterotic Line Bundle Study}

\def\andname{\hspace*{-0.5em}}
\author{Andrei Constantin}
\email[Email address: ]{andrei.constantin@physics.ox.ac.uk}
\affiliation{School of Mathematics, University of Birmingham, Watson Building, Edgbaston, Birmingham B15 2TT, United Kingdom}
\affiliation{Rudolf Peierls Centre for Theoretical Physics, University of Oxford, Parks Road, Oxford OX1 3PU, United Kingdom}
\author{Lucas T.-Y.\ Leung}
\email[Email address: ]{lucas.leung@physics.ox.ac.uk}
\affiliation{Rudolf Peierls Centre for Theoretical Physics, University of Oxford, Parks Road, Oxford OX1 3PU, United Kingdom}
\author{Andre Lukas}
\email[Email address: ]{andre.lukas@physics.ox.ac.uk}
\affiliation{Rudolf Peierls Centre for Theoretical Physics, University of Oxford, Parks Road, Oxford OX1 3PU, United Kingdom}
\author{Luca A. Nutricati}
\email[Email address: ]{luca.nutricati@physics.ox.ac.uk}
\affiliation{Rudolf Peierls Centre for Theoretical Physics, University of Oxford, Parks Road, Oxford OX1 3PU, United Kingdom}

\begin{abstract}
Deriving the Yukawa couplings and the resulting fermion masses and mixing angles of the Standard Model (SM) from a more fundamental theory remains one of the central outstanding problems in theoretical high-energy physics. It has long been recognised that string theory provides a framework within which this question can, at least in principle, be addressed. While substantial progress has been made in studying flavour physics in string compactifications over the past few decades, a concrete string construction that reproduces the full set of observed SM flavour parameters remains unknown. Here, we take a significant step in this direction by identifying two explicit $E_8 \times E_8$ heterotic string models compactified on a Calabi--Yau threefold with abelian, holomorphic, and poly-stable vector bundles with an (MS)SM spectrum. Subject to reasonable assumptions about the moduli, we show that these models reproduce the correct values of the quark and charged lepton masses, as well as the quark mixing parameters, at some point in their moduli spaces. The resulting four-dimensional theories are $\mathcal{N}=1$ supersymmetric, contain no exotic fields and realise a $\mu$-term suppressed to the electroweak scale. While the issues of moduli stabilisation and supersymmetry breaking are not addressed here, our main result constitutes a proof of principle: there exist choices of topology and moduli within heterotic string compactifications which allow for an MSSM spectrum with the correct flavour parameters.
\end{abstract}

\maketitle

\newpage

\section{Introduction} \label{sec:intro}

It has been recognised for nearly four decades that  heterotic string compactifications on Calabi--Yau (CY) threefolds provide a natural framework for embedding realistic particle physics into string theory~\cite{CHSW1985}. Beyond unification, this approach offers a significant advantage over conventional effective field theory: in string theory, all low-energy quantities, including the free parameters of the Standard Model (SM), are, in principle, calculable from the geometry and topology of the compactification. This opens the door to deriving the SM from first principles. Before such calculations can be performed, however, one must first construct compactifications that exactly reproduce the SM gauge group and particle spectrum, with no additional exotic fields or unwanted low-energy interactions. Achieving this has proven to be highly non-trivial. Generic string compactifications produce large gauge groups and complex matter spectra. The problem  is exacerbated by the vast landscape of consistent four-dimensional string theory solutions and the technical challenges of identifying explicit compactifications that satisfy key requirements such as anomaly cancellation, supersymmetry and spectrum constraints. 

While decades of progress have brought us closer, an explicit construction that reproduces the exact SM spectrum with realistic flavour data has remained elusive. Recently, however, advances in mathematical methods, computational tools and search algorithms have enabled the systematic exploration of large classes of compactifications leading to many models with the correct (MS)SM spectrum.
Among the most promising and tractable avenues are compactifications of the $E_8 \times E_8$ heterotic string on CY threefolds with vector bundles built as direct sums of line bundles~\cite{Anderson:2011ns,Anderson:2012yf}. This framework gives rise to a large and well-controlled class of models amenable to large-scale scanning~\cite{Anderson:2013xka, He:2013ofa, Braun:2017feb, Constantin:2018xkj, Larfors:2020ugo, Abel:2023zwg} and explicit computational analysis. Many such constructions have been shown to yield the exact SM gauge group, three chiral families and no exotic matter. Crucially, progress in this area has been driven by the development of mathematical and algorithmic tools that allow for the efficient computation of particle spectra via sheaf cohomology~\cite{Hubsch:1992nu, cicypackage, cohomCalg:Implementation, cicytoolkit, Constantin:2018hvl,Klaewer:2018sfl, Larfors:2019sie,Brodie:2019dfx,Brodie:2020wkd,Brodie:2020fiq,Brodie:2021nit,Constantin:2024ulu} and the computation of physical Yukawa couplings in terms of the compactification data~\cite{Braun:2007sn, Douglas:2006rr, Douglas:2006hz, Anderson:2011ed, Anderson:2010ke, Blesneag:2015pvz, Blesneag:2016yag, Ashmore:2019wzb, Anderson:2020hux, Jejjala:2020wcc, Douglas:2020hpv,Larfors:2021pbb,Larfors:2022nep, Ashmore:2021rlc, Ashmore:2023ajy, Constantin:2024yxh, Butbaia:2024tje}. 

A key structural feature of heterotic line bundle models is the presence of additional $U(1)$ gauge symmetries originating from the internal gauge bundle. While the corresponding gauge bosons typically acquire large St\"uckelberg masses and decouple from the low-energy spectrum, the symmetries themselves persist as effective global selection rules in the four-dimensional theory. Both SM fields and SM gauge-singlet moduli carry non-trivial charges under these $U(1)$ symmetries, which strongly constrain the form of allowed operators in the effective theory.
When the moduli acquire vacuum expectation values (VEVs), they induce effective SM couplings. This mechanism, reminiscent of Froggatt--Nielsen models~\cite{FROGGATT1979277}, can generate hierarchical structures in the couplings in a way that arises naturally from the geometry of the compactification. A systematic study of how such $U(1)$ symmetries constrain the operator structure of effective field theories with features characteristic of heterotic line bundle models was recently discussed in~\cite{Constantin:2024yaz}. The present work builds on that foundation but is taking a top-down rather than a bottom-up point of view. Our goal here is to identify explicit $E_8\times E_8$ heterotic line bundle models that successfully reproduce the observed flavour structure of the SM. 

The effective couplings in heterotic line bundle models, such as the Yukawa matrices, are shaped by two key ingredients. First, the $U(1)$ selection rules inherited from the compactification restrict the structure of allowed operators, often requiring the insertion of additional moduli fields to ensure symmetry invariance. When these moduli acquire VEVs, they generate suppression factors in the effective couplings, as the relevant VEVs, measured in units of the compactification scale (typically close to the GUT scale), are subunitary. Second, each operator comes with an order-one coefficient that depends on the detailed geometry of the compactification, for example, through overlap integrals of internal wavefunctions. While these coefficients depend non-trivially on the (complex structure and K\"ahler) moduli, numerical studies indicate they typically lie within an order of magnitude of unity across the moduli space~\cite{Constantin:2024yxh}. 

Assuming the unfixed coefficients vary within a few orders of magnitude of one, and that the moduli acquire generic, non-hierarchical VEVs, we identify two concrete line bundle models that reproduce the observed quark and charged lepton masses as well as the CKM matrix. 
The analysis is carried out in the context of the Minimal Supersymmetric Standard Model (MSSM). While we do not address how the moduli are stabilised at the required values, nor the mechanism of supersymmetry breaking, this approach is nonetheless well-motivated: in many string scenarios, the leading-order flavour structure is fixed in the supersymmetric limit, while supersymmetry breaking primarily affects soft terms and the scalar potential. As long as the moduli are stabilised near the VEVs identified in our analysis, the fermion masses and mixing angles remain essentially unchanged. Supersymmetry also imposes strong structural constraints on the effective theory: holomorphic quantities like superpotential Yukawa couplings are determined at tree level and protected from large quantum corrections, while the K\"ahler potential, although more intricate, is still computable in terms of the underlying geometry~\cite{Constantin:2024yxh}. These features make the computation of physical Yukawa couplings in terms of moduli-dependent suppression factors and order-one geometric coefficients, feasible.

In addition to reproducing the observed flavour structure, the models presented below allow for a $\mu$-term of the correct electroweak scale. The $\mu$-problem (that is, the question of why a supersymmetric Higgs mass parameter should lie far below the Planck or compactification scale) is resolved here by the same mechanism that controls the Yukawa hierarchies. The bare $\mu$-term is forbidden by the compactification topology, but an effective term is generated through higher-dimensional operators involving moduli fields. Once these moduli acquire appropriately suppressed VEVs, an effective $\mu$-term of the right magnitude is induced in our models.

\section{Heterotic Line Bundle Models}
{\bfseries Geometry.} The systematic study of heterotic line bundle models was initiated in Refs.~\cite{Anderson:2011ns,Anderson:2012yf, Anderson:2013xka, He:2013ofa}. These models are based on compactifying the $E_8 \times E_8$ heterotic string on smooth CY threefold quotients $X/\Gamma$,
where $X$ is a simply-connected CY threefold  with Picard number $h=h^{1,1}(X)$ and a freely-acting finite symmetry group~$\Gamma$. For the models presented here, the covering space $X$ is  a complete intersection of hypersurfaces in a product of projective spaces (a CICY).
Gauge symmetry breaking from the 10D $E_8 \times E_8$ theory to the 4D SM is achieved by specifying a non-trivial gauge background on the internal CY manifold in one of the $E_8$ sectors, referred to as the observable sector. This internal gauge field background  corresponds to a holomorphic vector bundle $V$ on $X$. We take $V$ to be a direct sum of five line bundles over  $X$,
\begin{equation}\label{Vdef}
V = \bigoplus_{a=1}^5 \mathcal{L}_a = \bigoplus_{a=1}^5 \mathcal{O}_X(\mathbf{k}_a)~,
\end{equation}
where the $h$-dimensional integer vectors $\mathbf k_a=(k_a^i)$ specify the first Chern classes $c_1(\mathcal{L}_a)$ and label the line bundles~$\mathcal{L}_a$. The line bundle sum $V$ can often be chosen so that it descends to a line bundle sum $\hat V = \oplus_{a=1}^5 \hat{\mathcal L}_a$ on  $X/\Gamma$ and this  is going to be the case for our examples. The bundle $\hat{V}$~is also required to be poly-stable somewhere inside the K\"ahler cone of $X/\Gamma$ and must satisfy the topological conditions $c_2(\hat V) \leq c_2(X/\Gamma)$. The latter condition ensures anomaly cancellation can be satisfied via the inclusion of five-branes wrapping holomorphic curves or a bundle in the other, hidden $E_8$ sector. To embed the structure group of $V$ into (the observable) $E_8$, we impose $c_1(V)=\sum_ac_1(\mathcal{L}_a)=0$ or, equivalently, $\sum_a\mathbf{k}_a=0$. Provided this condition is satisfied (and the line bundles are otherwise sufficiently generic) the structure group of~$V$ is  $\mathcal{G}=S(U(1)^5)\cong U(1)^4$. The resulting low-energy gauge group is $SU(5) \times \mathcal{G}$, the commutant of $\mathcal{G}$ within $E_8$. The $SU(5)$ factor can be interpreted as a GUT gauge group but it is further broken (and indeed never realised at the 4D field theory level) to the SM gauge group by a discrete Wilson line associated to the non-trivial fundamental group of the quotient manifold $X/\Gamma$. Hence, the final low-energy symmetry group (in the observable sector) is $G_{\rm SM}\times \mathcal{G}$, where $G_{\rm SM}= SU(3){\times} SU(2){\times} U(1)$ is the SM gauge group and the additional group $\mathcal{G}$ emerges as a flavour symmetry.
\vspace{4pt}

{\bfseries Spectrum.}
Although the physical 4D gauge group is precisely the SM gauge group $G_{\rm SM}$, the low-energy theory has features which mimic those of a conventional field theory GUT. This happened because $\mathcal{G}$ commutes with the underlying $SU(5)$ GUT symmetry and, as a result, all SM multiplets which originate from the same $SU(5)$ GUT multiplet must have the same $\mathcal{G}$ charge. This gives rise to a crude form of Yukawa unification: the allowed singlet insertions needed to form effective couplings are identical for, say, the $d$-quark and the charged lepton Yukawa couplings. However, the coefficients multiplying these operators---which arise from overlap integrals of internal wavefunctions---are typically different, and hence Yukawa unification is generically broken at the numerical level~\cite{Buchbinder:2016jqr}, in contrast to conventional GUTs.

It is important to note, however, that this is not the only possible situation in heterotic line bundle models. In certain constructions, such as the one discussed in Ref.~\cite{Buchbinder:2014qca} (which realises a KSVZ QCD axion with a decay constant in the phenomenologically allowed range), the MSSM spectrum at the abelian locus includes exotic vector-like states which can be lifted by giving suitable VEVs to $U(1)$-charged singlet fields. In such scenarios, once the moduli are displaced away from the abelian locus, the SM multiplets are no longer required to assemble into $SU(5)$ representations with a common $\mathcal{G}$ charge. This may lead to greater flexibility in the resulting effective low-energy Yukawa textures. A systematic exploration of such models lies beyond the scope of this work and is left for future investigation.

\begin{table}[!ht]
\begin{center}
	\begin{tabular}{|c|c|c||c|c|}
		\hline\hline
	\varstr{8pt}{4pt}		field & SM rep & name & SU(5) & $\cG$ charge pattern  \\[1pt]
		\hline
		\varstr{10pt}{4pt}	$Q$ & ~~$(\bm{3},\bm{2})_{1~~}$ & LH quark&$\bm{10}$ & $\uv_a$ \\[2pt]
		$u$ & ~~$(\bm{\bar{3}},\bm{1})_{-4}$& RH $u$-quark & & \\[2pt]
		$e$ & ~~$(\bm{1},\bm{1})_{6~~}$& RH electron& &  \\[2pt]\hline
	\varstr{10pt}{4pt}	$d$ & ~~$(\bm{\bar{3}},\bm{1})_{2~~}$ &RH $d$-quark& $\bm{\bar{5}}$ & $\uv_a+\uv_b$ \\[2pt]
		$L$ & ~~$(\bm{1},\bm{2})_{-3}$& LH lepton& & \\[2pt]\hline
	\varstr{10pt}{4pt}		$H^d$ & ~~$(\bm{1},\bm{2})_{-3}$&down-Higgs& $\bm{\bar{5}}^{H^d}$ & $\uv_a+\uv_b$ \\[2pt]
		$H^u$ & ~~$(\bm{1},\bm{2})_{3~~}$&up-Higgs& $\bm{5}^{H^u}$ & \!\!\!\!$-\uv_a-\uv_b$ \\[2pt]\hline
	\varstr{10pt}{4pt}		$\phi$ & $(\bm{1},\bm{1})_0$&pert.~moduli& $\bm{1}$ & $\uv_a-\uv_b$ \\[2pt]
		$\Phi^i$ & $(\bm{1},\bm{1})_0$&non-pert.~moduli & $\bm{1}$ & $( k_1^i,\ldots,k_5^i)$ \\[2pt]
		\hline
	\end{tabular}
\caption{The field content and multiplet charges of rank $5$ heterotic line bundle models, where ${\mathbf e}_1,\ldots ,{\mathbf e}_5$ are the standard five-dimensional unit vectors. The precise charge assignments depend on the topology of the underlying compactification.}\label{tab:fields} 
\vspace{-12pt}
\end{center}
\end{table}
While the group $\mathcal{G}$ is of rank four, charges are most naturally represented as five-component integer vectors $\bq \in \mathbb{Z}^5$, subject to the equivalence relation
\begin{equation*}
\bq \sim \bq' \quad \Leftrightarrow \quad \bq - \bq' \in \mathbb{Z}(1,1,1,1,1),
\label{equiv}
\end{equation*}
which reflects the trace-free condition for $\mathcal{G}=S(U(1)^5)$. The charge patterns of all relevant multiplets are summarised in Table~\ref{tab:fields}. Let us briefly explain their origin.

Suppose a quark doublet $Q$ arises as a zero mode of the Dirac operator on the quotient manifold $X/\Gamma$, valued in one of the five line bundles, say $\hat{\mathcal{L}}_a$, where $a \in {1,\ldots,5}$. Then $Q$ carries $\mathcal{G}$ charge ${\bf e}_a$, with $({\bf e}_a)_{a=1}^5$ denoting the standard five-dimensional unit vectors. In any given model, the three quark doublets may all descend from the same line bundle---leading to identical charges---or they may originate from different line bundles, resulting in distinct charges across generations. Similar considerations apply to right-handed up-quarks $u$ and electrons~$e$.
In contrast, the right-handed down-quarks $d$ and lepton doublets $L$ arise as zero modes associated with tensor products $\hat{\mathcal{L}}_a \otimes \hat{\mathcal{L}}_b$, and hence carry charges ${\bf e}_a + {\bf e}_b$. Again, the specific charge pattern across families is determined by where each field originates in the line bundle sum. As a general rule, in order to explain mass hierarchies, we require $\mathcal{G}$ to be a flavour symmetry, that is, not all families should have the same $\mathcal{G}$ charge. 
From the above discussion, this implies a topological constraint on the underlying string models: $(Q,u,e)$-families should be associated to at least two different line bundles, that is, we need $h^1(X/\Gamma,\hat{\mathcal{L}}_a)>0$ for more than one line bundle $\hat{\mathcal{L}}_a$.


The final two rows of Table~\ref{tab:fields} refer to moduli fields. The perturbative moduli $\phi$ arise as zero modes of $\hat{\mathcal{L}}_a \otimes \hat{\mathcal{L}}_b^*$ and carry charges ${\bf e}_a - {\bf e}_b$. A given compactification typically yields a spectrum of such singlets, which parameterise deformations of the vector bundle. When all $\phi$ VEVs vanish, the model sits precisely at the line bundle locus, while non-zero $\phi$ VEVs correspond to moving away from this locus towards non-abelian bundle configurations (see, e.g.~\cite{Buchbinder:2013dna, Buchbinder:2014sya} for a detailed case study).

Finally, consider the K\"ahler moduli $T^i$, $i=1,\ldots,h$, which transform under $\mathcal{G}$ as
$T^i \mapsto T^i + i \epsilon^a k_a^i$,
where $\epsilon^a$ are the $\mathcal{G}$ transformation parameters and $k_a^i$ are the integers specifying the line bundles in Eq.~\eqref{Vdef}. The non-perturbative moduli
$\Phi^i = \exp(-T^i)$
then carry $\mathcal{G}$ charges $(k_1^i, \ldots, k_5^i)$, as indicated in Table~\ref{tab:fields}.
\vspace{4pt}

{\bfseries Moduli VEVs.} SM operators are dressed with singlet insertions involving $\phi$ and $\Phi$ fields---representing perturbative and non-perturbative effects, respectively---in a manner controlled by $\mathcal{G}$ invariance. This mechanism provides a string-theoretic realisation of Froggatt--Nielsen models. Once these singlets acquire VEVs, effective couplings, including Yukawa couplings, are generated. The detailed structure of these couplings depends on the $\mathcal{G}$ charges of the matter fields, Higgs fields and moduli, which, as we have seen, are determined by the topology and geometry of the compactification.
Relying on this mechanism, we identify two explicit line bundle models that successfully reproduce the observed pattern of fermion masses and mixing angles of the SM. In doing so, we assume that the singlet VEVs and the order-one coefficients multiplying the operators can be chosen within plausible ranges, as detailed in Appendix~\ref{sec:SearchStrategy}. 

To be more precise, we reproduce the measured (infrared) values for the SM fermion masses and mixing angles in the following sense. While the VEVs for the perturbative and non-perturbative fields $\phi$ and $\Phi$ can be taken at face value, the order-one coefficients depend in a complex way on the underlying complex structure and K\"ahler moduli. In our analysis, these coefficients correspond to their infrared values. To connect these to a specific point in moduli space, one would need to (1)~run the order-one coefficients up to the compactification scale, taking into account the details of supersymmetry breaking, and (2) determine the map between the geometrical moduli and the order-one coefficients, as can be done using the tools developed in Ref.~\cite{Constantin:2024yxh}. The key point is that while this matching is technically involved, it is, in principle, feasible. What we take as order-one in the IR will differ from the corresponding UV values by factors of order unity arising from renormalization group evolution and the mapping to the underlying geometry. 

\vspace{4pt}

{\bfseries Couplings.} Having outlined how the observed fermion mass and mixing parameters can be reproduced in our framework, we now turn to the structure of the low-energy effective action that underlies these results. The dynamics of heterotic line bundle models at low energies is governed by a four-dimensional $\mathcal{N}=1$ supergravity theory with superpotential
\begin{equation*}
\begin{aligned}
W & = W_Y + W_R + \ldots \\
W_Y & = \mu H^u H^d {+} Y^u_{IJ} H^u Q^I u^J {+} Y^d_{IJ} H^d Q^I d^J {+} Y^e_{IJ} H^d e^I L^J \\
W_R & \supset \{L L e,~Q L d,~u d d,~L H^u\}
\end{aligned}
\end{equation*}
where $I, J = 1,2,3$ are family indices, $Y_{IJ}^u$, $Y_{IJ}^d$ and $Y_{IJ}^e$ denote the effective Yukawa couplings, and $\mu$ is an effective supersymmetric mass term. All these couplings are holomorphic functions of the moduli fields. The $\mu$-term and typically many Yukawa couplings are absent at the renormalisable level and arise effectively through insertions of singlet fields $\phi$ and $\Phi$, whose VEVs spontaneously break~$\mathcal{G}$.

In addition to the abelian symmetry~$\mathcal{G}$, we assume an R-parity symmetry that forbids all renormalisable R-parity violating terms in the superpotential, $W_R$. Crucially, we also assume that R-parity remains unbroken throughout moduli space, even when $\mathcal{G}$ is spontaneously broken, ensuring the absence of dangerous effective R-parity violating operators. Geometrically, R-parity can originate from a discrete symmetry of the quotient CY threefold $X/\Gamma$. The ellipsis in $W$ represents higher-dimensional operators, including those contributing to proton decay, which are naturally suppressed by the compactification scale.

The K\"ahler potential takes the standard form in $\mathcal{N}=1$ supergravity, depending on the K\"ahler moduli, complex structure moduli, and matter fields. The moduli dependence of the matter field K\"ahler metric leads to non-trivial wavefunction normalisations and kinetic mixing, modifying the effective Yukawa couplings upon canonical normalisation. However, following the perspective of Ref.~\cite{Constantin:2024yxh}, we regard these effects as order-one and absorb them into the order-one coefficients multiplying the superpotential operators. Similarly, any indirect impact of the gauge kinetic function~$f$, for example via RG evolution, can be absorbed into these coefficients.

\section{Results}
{\bfseries Geometry.}
Out of the 202 line bundle models constructed in Ref.~\cite{Anderson:2011ns}, we identified two that reproduce the observed quark and charged lepton masses and CKM matrix. Our search strategy is described in Appendix~\ref{sec:SearchStrategy}. The models are based on the CICY threefold:
\begin{equation*}
\begin{aligned}
X_{6770} &~\in~~ \cicy{
\IP^1 \\[2pt]
\IP^1 \\[2pt]
\IP^1 \\[2pt]
\IP^1 \\[2pt]
\IP^1
}{
\,1~ & 1~ \\[2pt]
1~ & 1~ \\[2pt]
1~ & 1~ \\[2pt]
2~ & 0~ \\[2pt]
0~ & 2~
}^{5,37} 
\end{aligned}
\end{equation*}
where the superscripts on the configuration matrix indicate the non-trivial Hodge numbers. The manifold is realised as a complete intersection of two hypersurfaces $\{p_1=0\}\cap\{p_2=0\}\subseteq(\IP^1)^{\times 5}$, where $p_1$, $p_2$ are homogeneous polynomials with multi-degrees given by the columns of the above matrix. To describe the relevant discrete symmetry~\cite{Braun:2010vc}, we introduce the notation $[x_{i,0}:x_{i,1}]$ for the homogeneous coordinates on the $i$-th $\IP^1$ factor. 
The manifold $X_{6770}$ admits a free action of $\Gamma=\mathbb{Z}_2$ defined by  $x_{i,0}\rightarrow -x_{i,0}$, $x_{i,1}\rightarrow x_{i,1}$, $p_1\rightarrow p_1$, $p_2\rightarrow -p_2$. The resulting quotient $X_{6770}/\mathbb Z_2$ has Hodge numbers $(5,21)$. 

The line bundle sums for the two models are given by
\begin{equation*}
\begin{aligned}
V_1=\left(\begin{array}{ccccc} ~1& ~0& ~0& ~0& \!\!-1\\[2pt] \!\!-2& ~1& ~1& ~0& ~0\\[2pt] ~0& ~1& ~0&\!\!-1& ~0\\[2pt] ~0&\!\!-2& ~0& ~1& ~1\\[2pt] ~0& ~0& \!\!-1& ~1& ~0 \end{array} \right), ~
V_2=\left(\begin{array}{ccccc} ~1& ~0& ~0& ~0& \!\!-1\\[2pt] \!\!-2& ~1& ~1& ~0& ~0\\[2pt] ~0& ~0&\!\!-1& ~0& ~1\\[2pt] ~1&\!\!-2& ~0& ~1& ~0\\[2pt] ~0& ~0& ~1& \!\!-1& ~0 \end{array} \right)
\end{aligned}
\end{equation*}
where the columns define the first Chern class vectors ${\mathbf k}_a$.
In both cases, all line bundles are individually equivariant with respect to the relevant freely acting symmetry. The choice of equivariant structure amounts to a choice of one $\Gamma$-representation for each line bundle. For our examples, all line bundles are associated with the trivial representation, except for the second line bundle in $V_1$ and the fifth line bundle in $V_2$. 
These choices are correlated with the Wilson line, which is introduced to project out the Higgs triplets while breaking $SU(5)$ to~$G_{\rm SM}$. The Wilson line is specified by two one-dimensional $\Gamma$-representations, $W_2$ and $W_3$, with $W_2 \neq W_3$. Specifically, we choose $(W_2,W_3)=(1,0)$ for both models.
Crucially, the four gauge bosons associated with the symmetry $\mathcal{G} = S(U(1)^5)$ acquire masses at the compactification scale via the Green--Schwarz mechanism and do not give rise to low-energy fifth forces.

We now present the low-energy physics associated with the two models. In both cases, the massless spectra consist of three chiral families of quarks and leptons, along with a single vector-like pair of Higgs doublets and several singlet fields. The $\mathcal G$ charges and the number of singlet fields are model-dependent. The perturbative parts of the spectra are determined through computations of the cohomologies $H^1(X,V)$, $H^1(X,\wedge^2 V)$, $H^1(X,\wedge^2 V^*)$ and $H^1(X,V\otimes V^*)$ which split into $\Gamma$-representations. The parts that survive the $SU(5)$ breaking are determined by projecting along the directions indicated by the Wilson line~\cite{Anderson:2012yf}. For the non-perturbative fields, we have checked that the GV invariants associated with the generators of the Mori cone of $X_{6770}$ are non-vanishing. However, we have not attempted to compute the corresponding Pfaffians and our implicit assumption is that these are non-vanishing.
\vspace{4pt}

{\bfseries Model 1 on $X_{6770}/\mathbb Z_2$.} The low-energy spectrum includes the following fields:

    \begin{table}[H]
    \centering
    \begin{tabular}{c|c}
    {\it Matter } & $(Q,u,e)_{\mathbf{e}_1},\ 2\,(Q,u,e)_{\mathbf{e}_2},$ \\[2pt]
     & $(d, L)_{\mathbf{e}_1+\mathbf{e}_2},\ (d, L)_{\mathbf{e}_1+\mathbf{e}_5},\ (d, L)_{\mathbf{e}_2+\mathbf{e}_5}$ \\[2pt]\hline
    \varstr{10pt}{4pt} {\it Higgs }  & $H^u_{-\mathbf{e}_1-\mathbf{e}_2},\ H^d_{\mathbf{e}_1+\mathbf{e}_2}$ \\[2pt]\hline
    \varstr{10pt}{4pt} &~~~ $\phi_{1,2}:=\phi_{\mathbf{e}_1-\mathbf{e}_2},\ ~~~\,4\,\phi_{2,1}:=4\,\phi_{\mathbf{e}_2-\mathbf{e}_1},\ $ \\[2pt]
    &~~~ $4\,\phi_{1,3}:=4\,\phi_{\mathbf{e}_1-\mathbf{e}_3},\ 3\,\phi_{5,1}:=3\,\phi_{\mathbf{e}_5-\mathbf{e}_1},\ $ \\[2pt]
    {\it Moduli}  &\, $~2\,\phi_{2,3}:=2\,\phi_{\mathbf{e}_2-\mathbf{e}_3},\ 7\,\phi_{2,5}:=7\,\phi_{\mathbf{e}_2-\mathbf{e}_5}$ \\[2pt]
    &~~ $\Phi_1:=\Phi_{(1,0,0,0,-1)},\ \Phi_2:=\Phi_{(-2,1,1,0,0)}$,\ \\[2pt]
    &~~ $\Phi_3:=\Phi_{(0, 1, 0, -1, 0)},\ \Phi_4:=\Phi_{(0, -2, 0, 1, 1)}$,\ \\[2pt] 
    &~~ $\Phi_5:=\Phi_{(0, 0, -1, 1, 0)}$ \\
    \end{tabular}
    \end{table}
The first step in assigning VEVs to the moduli fields is to ensure the correct scale for the $\mu$-term. We achieve this by setting
\begin{align*}
 \langle\Phi_3\rangle, \langle\phi_{2,1}\rangle,\langle\phi_{1,3}\rangle, 
            \langle\phi_{5,1}\rangle, \langle\phi_{2,3}\rangle = \mathcal{O}(M_{\text{EW}}/M_\text{GUT}) \,, 
\end{align*}
which guarantees that the $\mu$-term is of the electroweak scale. 
With these choices, the singlet insertions for the Yukawa matrices are given by the rank-2 matrices
\begin{equation*}
\begin{aligned}
\Lambda^u &=\left(
\begin{array}{ccc}
 0 & 1 & 1 \\
 1 & \phi _{1,2} & \phi _{1,2} \\
 1 & \phi _{1,2} & \phi _{1,2} \\
\end{array}
\right)\\
\Lambda^d_{11} &= \Phi_2 \Phi_4 \\
\Lambda^d_{12} &= \Phi_2^2 \Phi_5 \phi_{1,2}^2 + \Phi_2 \Phi_4 \phi_{2,5} \\
\Lambda^d_{13} &= \Phi_2^2 \Phi_5 \phi_{1,2}^3 + \Phi_2 \Phi_4 \phi_{2,5} \phi_{1,2} + \Phi_1 \Phi_2 \Phi_4 \\
\Lambda^d_{21} &= \Lambda^d_{31}= \Phi_2 \Phi_4 \phi_{1,2} \\
\Lambda^d_{22} &= \Lambda^d_{32}=\Phi_2^2 \Phi_5 \phi_{1,2}^3 + \Phi_2 \Phi_4 \phi_{2,5} \phi_{1,2} + \Phi_1 \Phi_2 \Phi_4 \\
\Lambda^d_{23} &= \Lambda^d_{33} = \Phi_2^2 \Phi_5 \phi_{1,2}^4 + \Phi_2 \Phi_4 \phi_{2,5} \phi_{1,2}^2 + \Phi_1 \Phi_2 \Phi_4 \phi_{1,2} 
\end{aligned}
\end{equation*}
omitting all terms of order $\mathcal{O}(M_{\text{EW}}/M_\text{GUT})$. Each entry in these matrices---more precisely, each individual term contributing to an entry when it is a sum of multiple terms---is multiplied by an order-one coefficient to produce the effective Yukawa matrices $Y^u_{IJ}=c^u_{IJ}\cdot\Lambda^u_{IJ}$, $Y^d_{IJ}=c^d_{IJ}\cdot\Lambda^d_{IJ}$, and $Y^e_{IJ}=c^e_{IJ}\cdot\Lambda^d_{IJ}$ (no summation). As discussed earlier, $Y^d_{IJ}$ and $Y^e_{IJ}$ are treated as independent even at the GUT scale. There is a large degeneracy among the possible choices of order-one coefficients and remaining VEVs that reproduce the observed masses and mixings in the IR. 
For illustration, we set 
\begin{align*}
\langle\phi _{1,2}\rangle{=}\langle\phi_{2,5}\rangle{=} 0.7\,,~~ 
\langle\Phi_1\rangle{=}\langle\Phi_2\rangle{=}\langle\Phi_4\rangle{=}\langle\Phi_5\rangle{=} 0.07\,.
\end{align*}
For the order-one coefficients in the up Yukawa matrix, one possible assignment is
\begin{equation*}
\begin{aligned}
c^u_{12} &= -0.2117, & c^u_{13} &= -0.1764, \\
c^u_{21} &= ~~0.2599, & c^u_{22} &= ~~3.0096, & c^u_{23} &= ~~2.4959, \\
c^u_{31} &= -0.2061, & c^u_{32} &= -2.3937, & c^u_{33} &= -1.9841.
\end{aligned}
\end{equation*}
For the down and lepton Yukawa matrices, there are multiple insertions contributing to each entry. To illustrate a simple choice, we assume that all such insertions contributing to a given entry are multiplied by a common order-one coefficient: 
\begin{equation*}
\begin{aligned}
\begin{aligned}
c^d_{11} &= -1.5947, & c^d_{12} &= ~~0.1109, & c^d_{13} &= ~~0.4794, \\
c^d_{21} &= ~~5.3584,  & c^d_{22} &= ~~0.1450, & c^d_{23} &= -1.5017, \\
c^d_{31} &= -4.2375, & c^d_{32} &= -0.1191, & c^d_{33} &= ~~1.1682.\\[4pt]
c^e_{11} &= ~~0.8238, & c^e_{12} &= ~~1.2625, & c^e_{13} &= ~~0.8958, \\
c^e_{21} &= -0.8783, & c^e_{22} &= -1.2210, & c^e_{23} &= -1.1892, \\
c^e_{31} &= ~~1.0613, & c^e_{32} &= ~~1.5147, & c^e_{33} &= ~~1.6658.
\end{aligned}
\end{aligned}
\end{equation*}
It is straightforward to verify that these choices, together with the choice 
\begin{equation*}
(\langle H^u\rangle, \langle H^d \rangle) = (48.930, 166.979)\, {\rm GeV}
\end{equation*}
for the Higgs VEVs reproduce the observed quark and charged lepton masses, as well as the CKM matrix. While many alternative choices are possible, the above values serve as a concrete proof of principle. 

This model also admits a second, qualitatively distinct mechanism for generating an electroweak-scale $\mu$-term. The key observation is that by choosing
\[
\begin{aligned}
& \langle\Phi_1\rangle, \langle\phi_{5,1}\rangle, \langle\phi_{2,5}\rangle = \mathcal{O}\bigl(\sqrt{M_{\text{EW}}/M_\text{GUT}}\bigr)\,, \\    
& \langle\Phi_3\rangle, \langle\phi_{2,1}\rangle, \langle\phi_{1,3}\rangle, \langle\phi_{2,3}\rangle = \mathcal{O}(M_{\text{EW}}/M_\text{GUT})\,,
\end{aligned}
\]
all singlet insertions allowed in the $\mu$-term are suppressed by at least a factor of $M_{\text{EW}}/M_\text{GUT}$. This effectively introduces an intermediate scale, reminiscent of the Kim–Nilles mechanism~\cite{Kim:1983dt} for solving the $\mu$-problem. In this scenario, the observed fermion masses and mixing parameters can still be reproduced with small adjustments to the order-one coefficients presented above, as can be verified by direct calculation. 
\vspace{8pt}

{\bfseries Model 2 on $X_{6770}/\mathbb Z_2$.} The low-energy spectrum includes the following fields:
    \begin{table}[H]
    \centering
    \begin{tabular}{c|c}
    {\it Matter } & $2\,(Q,u,e)_{\mathbf{e}_1},\ \,(Q,u,e)_{\mathbf{e}_2},$ \\[2pt]
     & $2\,(d, L)_{\mathbf{e}_1+\mathbf{e}_5},\ (d, L)_{\mathbf{e}_2+\mathbf{e}_5}$ \\[2pt]\hline
    \varstr{10pt}{4pt} {\it Higgs }  & $H^u_{-\mathbf{e}_2-\mathbf{e}_5},\ H^d_{\mathbf{e}_2+\mathbf{e}_5}$ \\[2pt]\hline
    \varstr{10pt}{4pt} &~~~ $8\,\phi_{1,2}:=8\,\phi_{\mathbf{e}_1-\mathbf{e}_2},\ \,6\,\phi_{1,3}:=6\,\phi_{\mathbf{e}_1-\mathbf{e}_3},\ $ \\[2pt]
    &~~~ $2\,\phi_{1,4}:=2\,\phi_{\mathbf{e}_1-\mathbf{e}_4},\ 4\,\phi_{5,1}:=4\,\phi_{\mathbf{e}_5-\mathbf{e}_1},\ $ \\[2pt]
    {\it Moduli}  &\, $~4\,\phi_{2,4}:=4\,\phi_{\mathbf{e}_2-\mathbf{e}_4},\ \phi_{2,5}:=\phi_{\mathbf{e}_2-\mathbf{e}_5}$ \\[2pt]
    &~~~ $6\,\phi_{3,5}:=6\,\phi_{\mathbf{e}_3-\mathbf{e}_5},\ $ \\[2pt]
    &~~ $\Phi_1:=\Phi_{(1,0,0,0,-1)},\ \Phi_2:=\Phi_{(-2,1,1,0,0)}$,\ \\[2pt]
    &~~ $\Phi_3:=\Phi_{(0, 0, -1, 0, 1)},\ \Phi_4:=\Phi_{(1, -2, 0, 1, 0)}$,\ \\[2pt] 
    &~~ $\Phi_5:=\Phi_{(0, 0, 1, -1, 0)}$ \\
    \end{tabular}
    \label{tab:model2}
    \end{table}

To obtain the correct scale for the $\mu$-term, we assign
\begin{align*}
 \langle\Phi_1\rangle, \langle\Phi_3\rangle, \langle\phi_{1,2}\rangle,\langle\phi_{1,3}\rangle, 
            \langle\phi_{1,4}\rangle, \langle\phi_{2,4}\rangle = \mathcal{O}(M_{\text{EW}}/M_\text{GUT}) \,, 
\end{align*}
With these choices, the singlet insertions for the Yukawa matrices are given by the rank-2 matrices
\begin{equation*}
\begin{aligned}
&~\Lambda^u =\left(
\begin{array}{ccc}
 \phi_{2,5}\,\phi_{5,1}^2 & \phi_{2,5}\,\phi_{5,1}^2 & \phi_{5,1} \\
 \phi_{2,5}\,\phi_{5,1}^2 & \phi_{2,5}\,\phi_{5,1}^2 & \phi_{5,1} \\
 \phi_{5,1} & \phi_{5,1} & 0 \\
\end{array}
\right)\\[4pt]
&\Lambda^d_{11} = \Lambda^d_{12} = \Lambda^d_{21} = \Lambda^d_{22}= \Phi_2 \Phi_4 \phi_{2,5} + \Phi_4 \phi_{2,5}^2 \phi_{3,5} \phi_{5,1}^2 \\
&\Lambda^d_{13} = \Lambda^d_{23} = \Lambda^d_{31} = \Lambda^d_{32} = \Phi_4 \phi_{2,5} \phi_{3,5} \phi_{5,1} \\
&\Lambda^d_{33} = \Phi_4\phi_{3,5} + \Phi_4^2 \Phi_5 \phi_{2,5}^2 \phi_{5,1} 
\end{aligned}
\end{equation*}
where we have omitted all terms of order $\mathcal{O}(M_{\text{EW}}/M_\text{GUT})$. Analogous to the model discussed above, there is significant degeneracy among the possible choices of order-one coefficients and the remaining VEVs that reproduce the observed masses and mixings in the IR.
For illustration, we display the following VEV choices:
\begin{align*}
&\langle\phi _{5,1}\rangle = \langle\phi_{2,5}\rangle = 0.7\,,\,\langle\phi_{3,5}\rangle = 0.5\, , \\[2pt]
&\langle\Phi_2\rangle = 0.07,~\langle\Phi_4\rangle = 0.01 ,~\langle\Phi_5\rangle = 0.06.
\end{align*}
For the order-one coefficients in the up Yukawa matrix, one possible assignment is
\begin{equation*}
\begin{aligned}
c^u_{11} &= -1.8576, & c^u_{12} &= \,\,~\,2.6696, & c^u_{13} &= 0.1381, \\
c^u_{21} &= -2.8519, & c^u_{22} &= \,\,~\,4.1054, & c^u_{23} &= 0.2093, \\
c^u_{31} &= \,\,~\,0.1084, & c^u_{32} &= -0.1776.
\end{aligned}
\end{equation*}
For the down-type quark and charged lepton Yukawa matrices, multiple insertions contribute to each entry. As before, for simplicity, we assume that all such insertions contributing to a given entry are multiplied by a single, common order-one coefficient: 
\begin{equation*}
\begin{aligned}
\begin{aligned}
c^d_{11} &= ~~1.2406, & c^d_{12} &= ~~2.5324, & c^d_{13} &= -5.3256, \\
c^d_{21} &= ~~1.9296,  & c^d_{22} &= ~~4.0020, & c^d_{23} &= -8.4645, \\
c^d_{31} &= -0.4503, & c^d_{32} &= -0.8237, & c^d_{33} &= ~~1.6341.\\[4pt]
c^e_{11} &= -2.8422, & c^e_{12} &= -1.9381, & c^e_{13} &= -2.3298, \\
c^e_{21} &= ~~2.9123, & c^e_{22} &= ~~2.0472, & c^e_{23} &=~~ 1.8753, \\
c^e_{31} &= -0.9057, & c^e_{32} &= -0.6317, & c^e_{33} &= -0.4543.
\end{aligned}
\end{aligned}
\end{equation*}
For the Higgs VEVs, we set
\begin{equation*}
(\langle H^u\rangle, \langle H^d \rangle) = (83.764, 152.511)\, {\rm GeV}\, .
\end{equation*}
It is straightforward to verify that these choices reproduce the observed quark and charged lepton masses, as well as the CKM matrix.
As in the previous model, here it is also possible to address the $\mu$-problem by introducing an intermediate scale of order $\sqrt{M_{\text{EW}}/M_{\text{GUT}}}$ and assigning the VEVs $\langle \phi_{3,5}\rangle$ and $\langle \Phi_3\rangle$ to this scale.
\vspace{8pt}

\section{Discussion and Conclusions}
\label{sec:conclusions}

In this work, we have presented explicit heterotic $E_8 \times E_8$ string compactifications on smooth Calabi--Yau threefolds with poly-stable, holomorphic, abelian vector bundles that yield the exact (MS)SM particle spectrum with realistic flavour data and no exotic states. Building on the framework of line bundle models developed over the past decade, we have exploited the residual global $U(1)$ selection rules inherited from the internal gauge bundle to generate hierarchical Yukawa couplings in the effective theory, in analogy to Froggatt--Nielsen models. By analysing these couplings under the constraints imposed by the $U(1)$ selection rules, we have identified explicit models capable of reproducing the observed quark and charged lepton masses and CKM mixing angles. While most singlet moduli were assumed to acquire generic, non-hierarchical VEVs, for each identified model we have determined the minimal subset that must be suppressed to generate an electroweak-scale $\mu$-term via higher-dimensional operators.

Our main result is a proof of principle: there exist explicit choices of CY topology and vector bundle data in heterotic string theory that yield a fully realistic (MS)SM spectrum with the correct flavour structure and no exotic states. The models rely on the existence of an R-parity symmetry---geometrically realised via a discrete symmetry of the quotient CY threefold---that forbids all renormalisable R-parity violating operators and remains unbroken throughout moduli space, ensuring the absence of dangerous low-energy couplings even after the spontaneous breaking of the $U(1)$s. The models identified here represent a significant step toward realising string compactifications that reproduce the SM in full detail.

While the number of free parameters in each model is significant---typically ranging from 50 to 80 due to the many moduli fields and order-one coefficients---this should not be mistaken for an unconstrained fit. The $U(1)$ selection rules inherited from the compactification sharply restrict the allowed operator structure, enforcing hierarchical Yukawa textures via higher-dimensional couplings involving $U(1)$-charged singlet moduli fields. These selection rules, along with the charges and multiplicities of the singlet fields themselves, are entirely dictated by the compactification data and not introduced ad hoc. The (infrared) values of the order-one coefficients are themselves constrained: they are expected to remain within a reasonable order-one range consistent with what can typically be computed from the compactification and subsequently evolved down to the infrared via renormalisation group running. In practice, we find that the vast majority of three-generation models in our scan fail to reproduce the observed flavour structure within these constraints, with only a small fraction successfully matching the measured masses and mixings. This indicates that the flavour structure arises non-trivially from the underlying compactification, rather than from arbitrary parameter tuning.

Of course, several open questions remain. Most notably, our analysis does not address the problem of moduli stabilisation. While we identify regions of moduli space consistent with the observed flavour data, we do not demonstrate that these regions can be dynamically realised through a concrete stabilisation mechanism. Additionally, the precise values of the order-one coefficients in the effective operators depend non-trivially on the CY geometry and can vary across moduli space. A complete UV-realisation would require computing these coefficients explicitly at a stabilised point using, for example, the neural network approach developed in Ref.~\cite{Constantin:2024yxh}.

Although we have not specified a stabilisation mechanism, our analysis implicitly relies on the assumption of low-energy supersymmetry, which ensures that holomorphic quantities such as the superpotential Yukawa couplings are protected from large quantum corrections, making their tree-level determination meaningful. Even though supersymmetry breaking and moduli stabilisation will generically shift the vacuum, one expects the leading-order flavour textures to remain robust provided the stabilised VEVs lie near the identified regions.

Future work should aim to embed these constructions into fully stabilised compactifications, incorporating fluxes or non-perturbative effects to fix the moduli dynamically while preserving the desired low-energy physics. It would also be valuable to extend the analysis to include neutrino masses and mixings, which could arise from similar higher-dimensional operators in the effective superpotential. 
Another promising direction is the systematic study of models in which exotic states at the abelian locus can be lifted via singlet VEVs, realising, for example, viable QCD axion scenarios with greater flexibility in the Yukawa structures.
Further line bundle models capable of realising the full range of SM flavour parameters can be identified by systematically searching the larger sets of three-generation models obtained in Refs.~\cite{Anderson:2013xka, He:2013ofa, Constantin:2018xkj, Abel:2023zwg}. The main lesson of this work is that the systematic exploration of large classes of heterotic compactifications, supported by appropriate computational geometry tools, offers a promising route to uncovering fully realistic string-derived models of particle physics.

\section*{Acknowledgements}
AC and LAN are supported by a Royal Society Dorothy Hodgkin Fellowship, grant DHF/R1/231142. LTYL is supported by the Oxford-Croucher Scholarship. AL acknowledges support by the STFC consolidated
grant ST/X000761/1. We would like to thank Jonathan Patterson for assistance with the Oxford Theoretical Physics computing cluster Hydra and Russell Jones for general technical support.

\bibliography{main}
\bibliographystyle{utcaps}

\appendix
\section{SM Yukawa couplings}
The quark and charged lepton mass matrices are related to the Yukawa matrices through the relations
\begin{equation}
    M^u=\langle H^u\rangle\, Y^u\;,~ M^d=\langle H^d\rangle\, Y^d\;,~ M^e=\langle H^d\rangle\, Y^e\;,
\end{equation}
where $\langle H^u\rangle$ and $\langle H^d\rangle$ satisfy
\begin{equation}
 \sqrt{\langle H^u\rangle^2+\langle H^d\rangle^2}\simeq 174\;{\rm GeV}\;.
\label{eq:UpDown_higgs_vevs}
\end{equation}
The quark and charged lepton masses are obtained from the singular value decomposition of the mass matrices: 
\begin{equation}\label{diag}
\begin{array}{rclcrcl}
 M^u &=& U^u \hat{M}^u (V^u)^\dagger&\qquad&\hat{M}^u &=& \text{diag}(m_u,m_c,m_t)\\
 M^d &=& U^d \hat{M}^d (V^d)^\dagger&&\hat{M}^d &=& \text{diag}(m_d,m_s,m_b)\\
  M^e &=& U^e \hat{M}^e (V^e)^\dagger&&\hat{M}^e &=& \text{diag}(m_e,m_\mu,m_\tau)\; .
 \end{array}
\end{equation} 
while the CKM quark mixing matrix is given by
\begin{equation}\label{VCKM}
        V_{\rm CKM} = (U^u)^\dagger U^d~.
\end{equation}
\begin{table}[H]
\renewcommand{\arraystretch}{1.35}
\centering
\begin{tabular}{|c|c|c|c|}
\hline
\textbf{Quark} & \(m_u\) (GeV) & \(m_c\) (GeV) & \(m_t\) (GeV) \\
\hline
\textbf{Mass} & \(0.00216^{+0.00049}_{-0.00026}\) & \(1.27 \pm 0.02\) & \(172.4 \pm 0.07\) \\
\hline
\textbf{Quark} & \(m_d\) (GeV) & \(m_s\) (GeV) & \(m_b\) (GeV) \\
\hline
\textbf{Mass} & \(0.00467^{+0.00048}_{-0.00017}\) & \(0.093^{+0.011}_{-0.005}\) & \(4.18^{+0.03}_{-0.02}\) \\
\hline\hline
\textbf{Lepton} & \(m_e\) (MeV) & \(m_\mu\) (GeV) & \(m_\tau\) (GeV) \\
\hline
\textbf{Mass} & \(0.511 \pm 0.001\) & \(0.1057 \pm 0.0001\) & \(1.77682 \pm 0.00016\) \\
\hline
\end{tabular}
\caption{\small Masses of quarks and charged leptons~\cite{ParticleDataGroup:2024cfk}.}
\label{tab:exp_mass}
\end{table}

\noindent For completeness, the standard reference values for the quark and lepton masses are listed in Table~\ref{tab:exp_mass}. We recall that the values for the charged lepton masses correspond to their physical (pole) masses, as these particles are directly observable. For quarks, which are confined, the masses are scheme- and scale-dependent parameters defined in a particular renormalisation prescription. The light-quark ($u$, $d$, $s$) masses are computed using dimensional regularisation in the minimal subtraction scheme  at a scale of $2\,\text{GeV}$, while the charm and bottom quark masses, $m_c(m_c)$ and $m_b(m_b)$, are given at their own mass scales in the same scheme. The top quark mass is conventionally quoted as its pole mass. 
\vspace{4pt}

\section{Search Strategy}
\label{sec:SearchStrategy}

The present search builds on the 202 line bundle models identified in Ref.~\cite{Anderson:2011ns} that yield three chiral families of fermions. After GUT breaking and computation of the graded cohomologies using the methods of Ref.~\cite{Constantin:2016xlj}, these models give rise to 2113 Standard Model-like spectra without exotic fields, in particular without Higgs triplets. Among these, 1568 models contain exactly one pair of Higgs doublets.  However, many of these models are physically equivalent in the sense that they share identical low-energy spectra and couplings, leaving 226 inequivalent models in total. This subset forms the starting point of our analysis.

For each of these models, we systematically compute the $\mathcal{G}$-invariant operators in the superpotential that contribute to the effective Yukawa couplings (including all terms with up to seven singlet insertions) and to the effective $\mu$-term (including terms with up to fifteen singlet insertions). The $\mu$-term must be suppressed to the electroweak scale---approximately fourteen orders of magnitude below the GUT scale---and we identify the minimal subsets of singlet VEVs that need to be suppressed by this factor in order to achieve the required scale while allowing the remaining VEVs to take generic values of order $0.01-0.1$. We then proceed as follows:

\begin{enumerate}
    \item Given the required suppression of certain singlet VEVs (by $\sim10^{-14}$), these VEVs are set to zero when computing the flavour parameters. This can reduce the rank of the Yukawa matrices. If both up- and down-type Yukawa matrices become rank one under this restriction, the model is discarded.

    \item The CKM matrix in Eq.~\eqref{VCKM} is enforced by randomly generating orthogonal matrices $U^u$ and $U^d$ until the product $(U^u)^\dagger U^d$ matches the target CKM matrix within a 10\% average deviation (in the absolute values of its entries). Note that fitting the CP-violating phase in the CKM matrix can be easily accomplished, since every singlet VEV and order-one coefficient comes with a phase. We, therefore, choose all VEVs and order-one coefficients to be real in order to keep the dimension of the parameter space manageable.

    \item For each realisation of the CKM matrix and each choice of suppressed singlet VEVs, there remain approximately 50–80 additional parameters that must be optimised to reproduce realistic quark and charged lepton masses. These include the Yukawa coefficients, the up-type Higgs VEV, the unsuppressed singlet VEVs and the entries of $V^u$, $V^d$, $U^e$ and $V^e$. The number of moduli fields and the Yukawa coefficients varies from model to model, with the latter depending on the number of leading operators appearing in the Yukawa matrices. We perform the optimisation of parameters using the trust-region reflective algorithm, a numerical method that enforces the parameter bounds specified below while minimising the loss function
    \begin{align}
        L ~=~ \sum_{\lambda = u,d,e} \left\| M^\lambda - U^\lambda \hat{M}^\lambda (V^\lambda)^\dagger \right\|_F^2 \,,
    \end{align}
    where $\|\cdot\|_F$ is the Frobenius norm.  Models for which $L < 10^{-4}$ are classified as “viable.” For each configuration of suppressed singlet VEVs consistent with solving the $\mu$-problem, we run the optimisation over several hundred random initialisation points and CKM matrix realisations. 
  
\end{enumerate}

The bounds imposed on the parameters in the optimisation are chosen to reflect physically reasonable values. The bundle moduli VEVs $\langle\phi_a\rangle$ are allowed to vary between 0.1 and 0.7, while the non-perturbative moduli VEVs $\langle\Phi_i\rangle$ range from 0.01 to 0.07. The order-one Yukawa coefficients $c^\lambda_{IJ}$ (for $\lambda = u,d,e$) are constrained as $0.1<|c^\lambda_{IJ}| < 9$. The up-type Higgs VEV $v_u$ is varied between 40GeV and 170GeV, consistent with electroweak symmetry breaking constraints. Finally, the twelve Euler angles used to parametrise the residual orthogonal matrices (after fixing $U^u$ and $U^d$) are left unconstrained.

\end{document}